# The Cambridge-Cambridge ROSAT Serendipity Survey – II. Classification of X-ray Luminous Galaxies

B. J. Boyle,[1] R. G. McMahon,[2] B. J. Wilkes,[3] Martin Elvis[3]

1. Royal Greenwich Observatory, Madingley Road, Cambridge, CB3 0EZ
2. Institute of Astronomy, University of Cambridge, Madingley Road, Cambridge, CB3 0HA
3. Center for Astrophysics, 60 Garden St, Cambridge, MA 02138, USA

**ABSTRACT**

We present the results of an intermediate-resolution (1.5Å) spectroscopic study of 17 X-ray luminous narrow emission-line galaxies previously identified in the Cambridge-Cambridge *ROSAT* Serendipity Survey and the *Einstein* Extended Medium Sensitivity Survey. Emission-line ratios reveal that the sample is composed of ten Seyfert and seven starburst galaxies. Measured linewidths for the narrow H$\alpha$ emission lines lie in the range $170 - 460 \,\mathrm{km\,s^{-1}}$. Five of the objects show clear evidence for asymmetry in the [OIII]$\lambda$5007 emission-line profile. Broad H$\alpha$ emission is detected in six of the Seyfert galaxies, which range in type from Seyfert 1.5 to 2. Broad H$\beta$ emission is only detected in one Seyfert galaxy. The mean full width at half maximum for the broad lines in the Seyfert galaxies is $\mathrm{FWHM} = 3900 \pm 1750 \,\mathrm{km\,s^{-1}}$. Broad ($\mathrm{FWHM} = 2200 \pm 600 \,\mathrm{km\,s^{-1}}$) H$\alpha$ emission is also detected in three of the starburst galaxies, which could originate from stellar winds or supernovae remnants. The mean Balmer decrement for the sample is H$\alpha$/H$\beta$ = 3, consistent with little or no reddening for the bulk of the sample. There is no evidence for any trend with X-ray luminosity in the ratio of starburst galaxies to Seyfert galaxies. Based on our previous observations, it is therefore likely that both classes of object comprise $\sim 10$ per cent of the 2 keV X-ray background.

## 1 INTRODUCTION

A number of recent spectroscopic surveys of soft (0.5–2 keV) X-ray sources detected at faint fluxes with the ROSAT mission (Boyle et al. 1995, Georgantopoulos et al. 1995) have all confirmed that, while QSOs comprise in excess of 50 per cent of the total X-ray population down to these flux levels, an increasingly large number of X-ray luminous, narrow ($\mathrm{FWHM} < 1000 \,\mathrm{km\,s^{-1}}$) emission-line galaxies (NLXGs) are identified as counterparts to faint X-rays sources with fluxes $S(0.5-2 \,\mathrm{keV}) < 10^{-13} \,\mathrm{erg\,s^{-1}\,cm^{-2}}$. These galaxies have X-ray luminosities in the range $10^{42} - 10^{43.5} \,\mathrm{erg\,s^{-1}}$, over 100 times more luminous than late-type galaxies (Fabbiano 1989), whose low-resolution optical spectra they most closely resemble. In a previous paper in this series (Boyle et al. 1995; hereinafter Paper I), we have demonstrated that, based on their space density and cosmological evolution, these emission-line galaxies could comprise between 15–30 per cent of the soft 0.5–2 keV X-ray background. This is entirely consistent with the upper limit of $\sim 50$ per cent for the contribution of QSOs to the 0.5–2 keV X-ray background, based on their luminosity function (Boyle et al. 1994), clustering properties (Georgantopoulos et al. 1993) and X-ray spectra (Georgantopoulos et al. 1995).

Unfortunately, due to the poor quality of many of the identification spectra, little is known about the precise nature of this population. In particular, it is not clear whether these emission-line galaxies are examples of starburst galaxies or 'hidden' active galactic nuclei (e.g. Seyfert 2 galaxies), both of which have previously been suggested as possible significant contributors to the X-ray background (Griffiths & Padovani 1990, Fabian & Barcons 1992) and are known to exist in X-ray surveys (e.g. Böller et al. 1992), albeit at much higher X-ray flux levels and lower space densities.

In order to understand the origin of this potentially significant population of X-ray sources, we report in this paper on a detailed intermediate-resolution spectroscopic study of 17 NLXGs, 10 of which have been identified in the Cambridge-Cambridge *ROSAT* Serendipity Survey (CRSS, see Paper I) and a further 7 objects selected at random from the *Einstein* Extended Medium Sensitivity Survey (EMSS, Stocke et al. 1991) which we suspect are similar (Paper I). This sample comprises all but two of the NLXG identified in the CRSS (CRSS1514.4+5627 and CRSS1605.9+2554), which were not observed due to lack of time.

In Section 2 we report on the observation and analysis of the NLXG spectra. Based on the results obtained from these spectra, we discuss the properties and classification of the NLXG in Section 3, including the implications for the composition for the soft X-ray background. We present our conclusions in Section 4.

**Table 1** Log of Observations

| Name | RA (J2000) h m s | Dec ° ′ ″ | z | L(0.5-2 keV) ($\times 10^{43}\,\mathrm{erg\,s^{-1}}$) | V* | Exposure (seconds) | Dichroic | Date |
|---|---|---|---|---|---|---|---|---|
| **CRSS NLXG** | | | | | | | | |
| CRSS0009.0+2041 | 00 09 01.7 | 20 41 36 | 0.189 | 0.34 | 19.0 | 3600 | – | 1993 Sept 20/21 |
| CRSS0030.2+2611 | 00 30 17.4 | 26 11 38 | 0.077 | 0.17 | 16.0 | 900 | 6100 | 1994 June 10/11 |
| CRSS0030.7+2616[1] | 00 30 47.9 | 26 16 50 | 0.246 | 0.59 | 18.4 | 3600 | – | 1993 Sept 20/21 |
| CRSS1406.7+2838 | 14 06 47.9 | 28 38 53 | 0.331 | 1.52 | 21.5 | 1800 | 7500 | 1994 June 10/11 |
| CRSS1412.5+4355 | 14 12 31.6 | 43 55 36 | 0.094 | 0.89 | 16.2 | 1800 | 6100 | 1994 June 9/10 |
| CRSS1413.3+4405 | 14 13 19.9 | 44 05 34 | 0.136 | 0.45 | 17.2 | 1800 | 6100 | 1994 June 9/10 |
| CRSS1415.0+4402 | 14 15 00.1 | 44 02 08 | 0.136 | 0.25 | 17.6 | 1800 | 6100 | 1994 June 9/10 |
| CRSS1429.0+0120 | 14 29 04.7 | 01 20 17 | 0.102 | 0.20 | 16.6 | 1800 | 6100 | 1994 June 9/10 |
| CRSS1605.6+2554 | 16 05 39.9 | 25 43 10 | 0.278 | 1.65 | 18.9 | 1800 | 7500 | 1994 June 10/11 |
| CRSS1705.3+6049[2] | 17 05 18.3 | 60 49 54 | 0.572 | 4.72 | 21.0 | 1800 | 6100 | 1994 June 10/11 |
| **EMSS NLXG** | | | | | | | | |
| MS1252.4−0457[3] | 12 54 58.1 | −05 13 22 | 0.158 | 1.47 | 19.1 | 1800 | 6100 | 1994 June 10/11 |
| MS1334.6+0351 | 13 37 09.8 | 03 35 55 | 0.136 | 1.36 | 17.7 | 900 | 6100 | 1994 June 9/10 |
| MS1412.8+1320 | 14 15 16.9 | 13 06 01 | 0.139 | 1.23 | 18.7 | 900 | 6100 | 1994 June 9/10 |
| MS1414.8−1247 | 14 17 34.0 | −13 01 06 | 0.198 | 7.42 | 19.3 | 1800 | 6100 | 1994 June 10/11 |
| MS1555.1+4522 | 15 56 40.6 | 45 13 38 | 0.181 | 3.80 | 18.0 | 1800 | 6100 | 1994 June 9/10 |
| MS1614.1+3239 | 16 16 01.9 | 32 32 25 | 0.118 | 0.96 | 17.0 | 1200 | 6100 | 1994 June 9/10 |
| MS2044.1+7532 | 20 43 18.3 | 75 43 39 | 0.183 | 2.08 | 20.1 | 900 | 6100 | 1994 June 9/10 |

\* O mag for CRSS NLXG.
[1] Declination incorrect in Paper I.
[2] X-ray luminosity incorrectly rounded-off in Paper I.
[3] Wrong source identified as $z = 0.158$ AGN in Stocke et al. (1991).

## 2  DATA

### 2.1  Observations

We obtained intermediate-resolution spectra of 17 emission-line galaxies previously identified in the CRSS and EMSS using the ISIS double arm spectrograph at the WHT on the nights of 1994 June 9/10 and 10/11. We operated ISIS with the Tektronix CCD on the blue arm and the EEV CCD on the red arm. We used 600 lines mm$^{-1}$ gratings in both arms, giving an instrumental resolution of 1.5 Å (0.67Å pix$^{-1}$). For each galaxy, we observed the redshifted H$\beta$/[OIII]$\lambda$5007 and H$\alpha$/[NII]$\lambda\lambda$6717,6731 regions in the blue and red arms respectively. Throughout the run, conditions were good (1 arcsec seeing) and we observed all objects with a 1.0 arcsec slit. The redshifts of the program objects allowed us to observe the H$\beta$/[OIII]$\lambda$5007 and H$\alpha$/[NII]$\lambda\lambda$6717,6731 regions in all NLXG with only one change of grating position and dichroic. Details of the observations, including exposure times for each program object, are given in table 1. A further two NLXG were observed with ISIS as part of the WHT service observation program on the night of 1993 September 20/21. The observations were made with the ISIS red arm, EEV detector and the 1200 lines mm$^{-1}$ grating, giving an overall resolution of 0.7 Å (0.33Å pix$^{-1}$) in the H$\alpha$/[NII]$\lambda\lambda$6717,6731 region. X-ray luminosities in the 0.5–2 keV band and optical magnitudes for all NLXG are also listed in table 1. The V magnitudes listed for the EMSS QSOs are taken from Stocke et al. (1991). The Palomar O plate magnitides for the CRSS were obtained from the APM Northern Sky Survey (Irwin, Maddox & McMahon 1994). The optical magnitudes correspond to the total galaxy magnitude and cannot therefore be used as an accurate measure of the nuclear magnitudes for these objects. To derive the X-ray luminosities we assumed $H_0 = 50\,\mathrm{km\,s^{-1}\,Mpc^{-1}}$, $q_0 = 0.5$ and an X-ray spectral slope $\alpha_X = 1$ ($f_X \propto \nu^{-\alpha_X}$). To convert the EMSS 0.3–3.5 keV band luminosities to the 0.5–2.5 keV band we divided by a factor of 1.8, the approximate conversion factor between the *Einstein* and *ROSAT* bands for a spectral index $\alpha_X = 1$ (see Boyle et al. 1995).

### 2.2  Data Reduction and Analysis

The data were reduced using standard routines in the IRAF reduction package on the Cambridge SPARC cluster. Optimally-extracted galaxy spectra were wavelength-calibrated using copper-argon arc spectra taken throughout each night during the observing run. The spectra were flux-calibrated using spectrophotometric standards taken during evening and morning twilight on each night. We present the reduced spectra for each galaxy in figure 1. The large gap in each spectrum corresponds to the gap in the wavelength coverage between the blue and red arms of ISIS.

Before measuring the emission lines, we first shifted each spectrum to the rest-frame using the redshift given in Table 1 and then divided each spectrum by a second-order polynomial fit to its own continuum, after excluding the regions ±150Å around the H$\beta$/[OIII] and H$\alpha$/[NII] lines. This division created a flat continuum (advantageous in the line-fitting routine described below) without removing any weak, broad features which may be present in the H$\alpha$ or H$\beta$ emission.

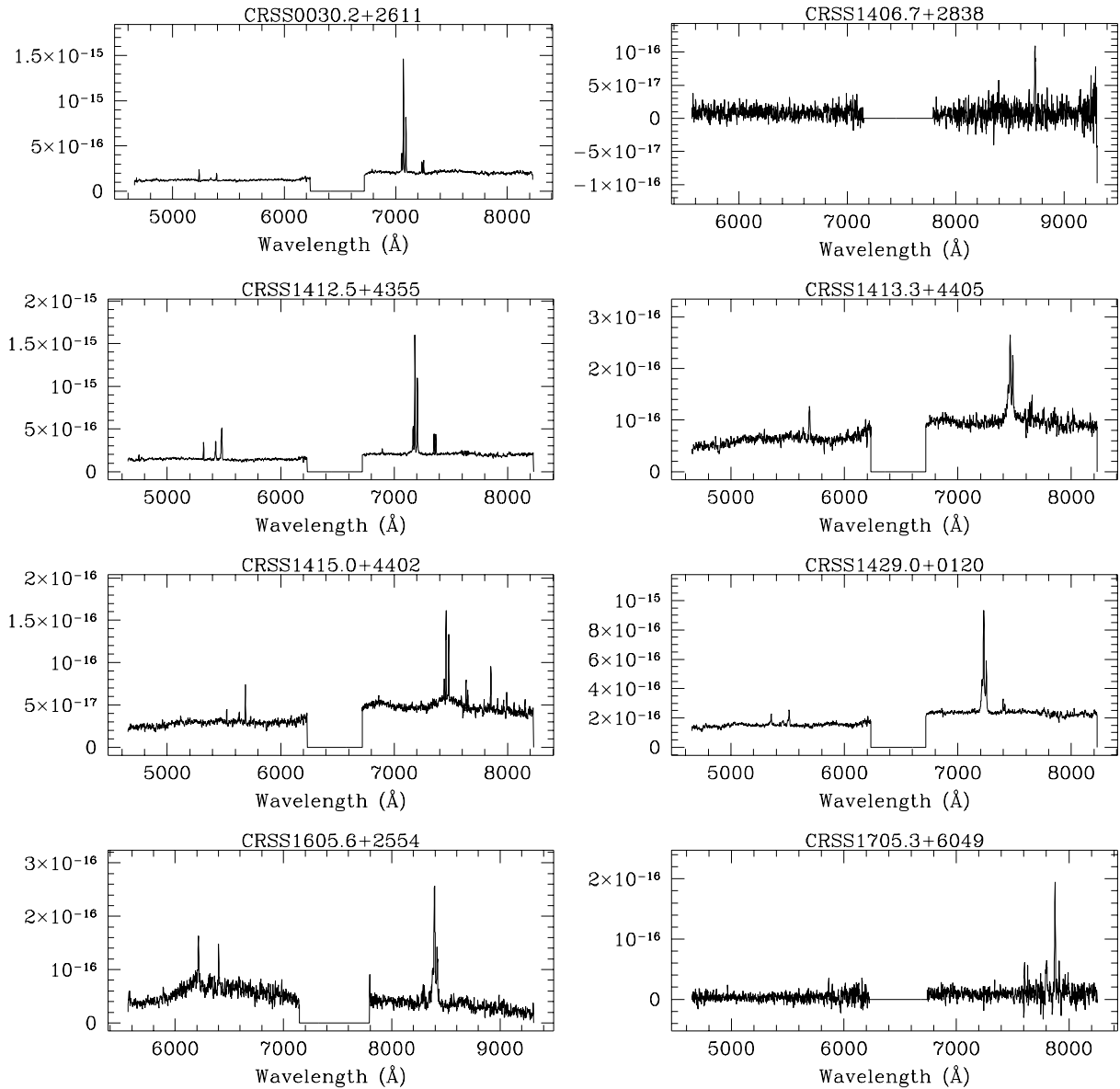

**Figure 1** Intermediate-resolution (1.5 Å) spectra for all 17 narrow emission-line galaxies observed in this program.

We then used the SPECFIT routine (written by Dr Gerard Kriss) in IRAF to measure emission line ratios, rest equivalent widths ($W$) and full width at half maximum intensity (FWHM) of the prominent emission lines in the spectra of the galaxies. This routine uses a Marquand $\chi^2$-minimisation routine to fit a user-specified number of functions (power-law or linear continuum, gaussian or logarithmic line profiles) to the input spectrum. In this case, the simplifying step of continuum division allowed us to fix the continuum at a constant value of 1. The fitting process yields typical Poission errors of 15 per cent and 10 per cent in the measurement of the equivalent widths and FWHM respectively.

For each of the 13 spectra in which [OIII]$\lambda$5007 was observed at a high signal-to-noise ratio (excluding CRSS1406.7+2838 and CRSS1705.3+6049, see Fig. 1), we first used the SPECFIT routine to fit both gaussian and logarithmic profiles to this emission line, in order to establish the correct profile shape to fit to the narrow emission lines. We chose the [OIII]$\lambda$5007 line for this purpose because it is the strongest narrow line observed in each spectrum which does not have any weak, broad (FWHM > 1000 km s$^{-1}$) component and is not blended with any other lines. We found that, in every case, the $\chi^2$ value for the gaussian profile fit was less than that for the logarithmic profile fit. In 10 cases, the $F$-ratio test (based on the ratio of the $\chi^2$ statistics, see Mood

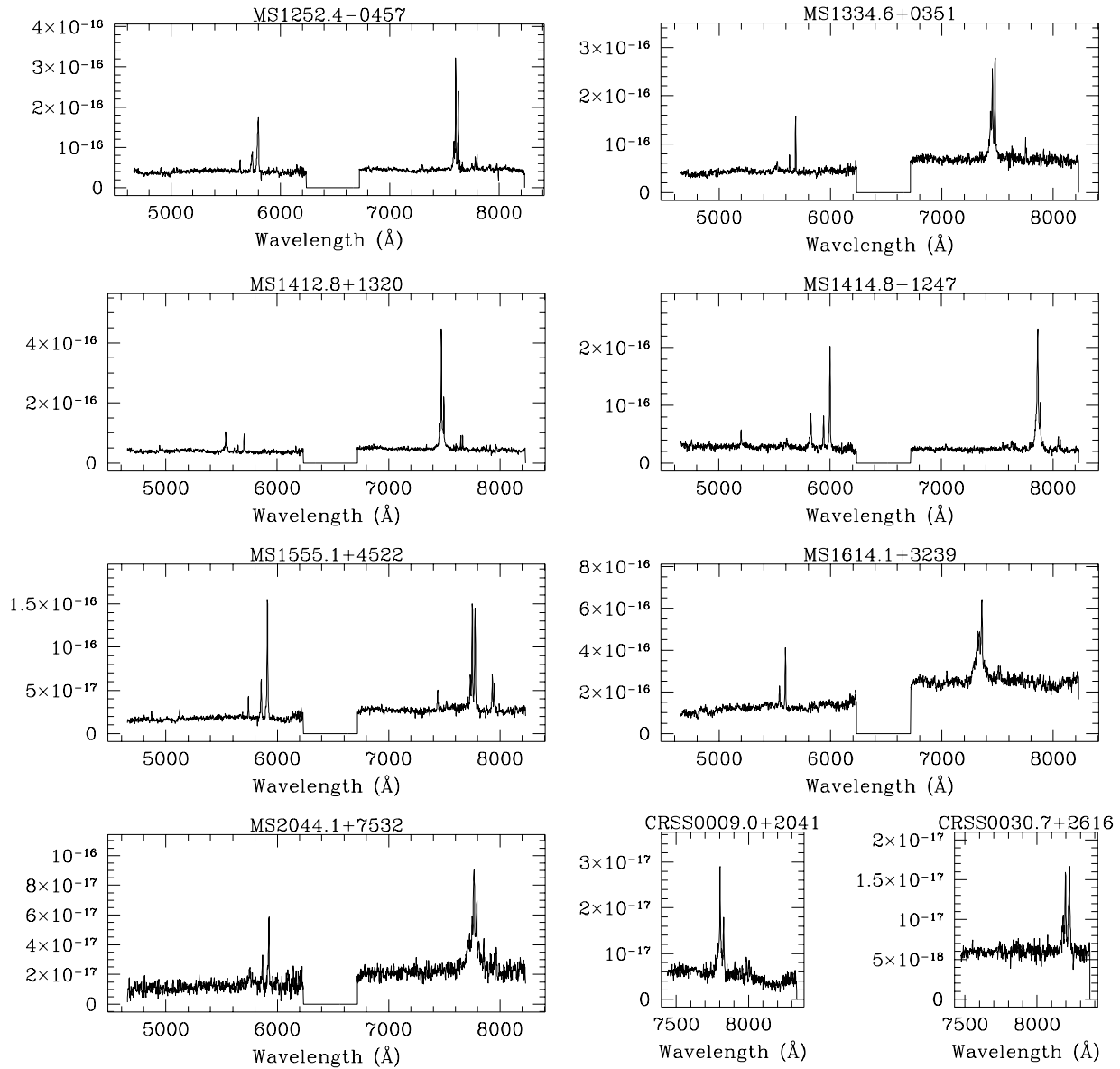

**Figure 1** contd.

& Graybill 1963) implied that the gaussian fit was preferred at the 99 per cent confidence level over the logarithmic fit. Based on this test, we subsequently used gaussian profiles throughout the fitting procedure.

Several of the [OIII]$\lambda$5007 emission lines appeared to exhibit a significant blue asymmetry. To quantify this observation, we performed another $F$-ratio test, this time on the $\chi^2$ values obtained from fitting a single and double gaussian to the [OIII]$\lambda$5007 line. In the latter case, the second gaussian component was blueshifted with respect to the rest wavelength of the line. We also investigated a double gaussian with a redshifted component. We stress that the use of this double gaussian fit is not intended to reflect any *physical* significance for the origin of any asymmetry. It simply provides us with a simple way to establish the significance of the asymmetry and obtain a more accurate measurement of the equivalent width, while retaining a consistent measurement process with all the other emission lines.

In five cases ($\sim$ 35 per cent of the sample) we found a significant blue asymmetry, with the extra gaussian component improving the preferred fit at the 99 per cent confidence level. No redshifted components were detected at the same level of significance. The NLXG in which the blueshifted components were observed are CRSS1429.0+0120, MS1252.4−0457, MS1414.8−1247, MS1555.1+4522 and MS2044.1+7532. Examples of these

**Table 2** Rest-frame emission line properties of NLXG sample

| Name | $W_{H\beta}$ (Å) | $W_{[OIII]}$ (Å) | [OIII] $a(20\%)^\dagger$ | $H\beta/[OIII]$ FWHM (km s$^{-1}$) | $W_{[OI]}$ (Å) | $W_{H\alpha}$ (Å) | $H\alpha/[NII]$ FWHM (km s$^{-1}$) | $W_{[NII]}$ (Å) | $W_{[SII]}$ (Å) | ID |
|---|---|---|---|---|---|---|---|---|---|---|
| **CRSS NLXG** | | | | | | | | | | |
| CRSS0009.0+2041 | | | | | 0.6 | 28.2 | 289 | 15.8 | 4.0 | 3.3 | Narrow |
| | | | | | | 8.1 | 358 | 8.6 | | | Blueshifted Hα/[NII]? |
| CRSS0030.2+2611 | 3.8 | 3.1 | 0.15 | 238 | 0.8 | 25.6 | 197 | 13.6 | 3.2 | 3.5 | Narrow |
| | | | | | | 8.0 | 2366 | | | | Broad Hα (weak) |
| CRSS0030.7+2629 | | | | | 1.0 | 12.5 | 421 | 16.0 | | | Narrow |
| CRSS1406.7+2838 | 8.0 | 4.9 | | 157 | – | 78.7 | 237 | 11.0 | 6.9 | 4.9 | Narrow |
| CRSS1412.5+4355 | 6.4 | 25.5 | 0.13 | 479 | 1.5 | 33.1 | 209 | 20.9 | 6.6 | 5.9 | Narrow |
| | | | | | | 20.4 | 3007 | | | | Broad Hα (weak) |
| CRSS1413.3+4405 | 1.7 | 9.1 | −0.08 | 491 | 1.4 | 9.6 | 338 | 7.0 | 1.1 | 2.6 | Narrow |
| | | | | | | 22.6 | 2353 | | | | Broad Hα |
| CRSS1415.0+4402 | 1.9 | 7.1 | 0.13 | 269 | 1.2 | 9.3 | 178 | 7.2 | 2.8 | 2.7 | Narrow |
| | | | | | | 49.0 | 7668 | | | | Broad Hα |
| CRSS1429.0+0120 | 4.2 | 5.7 | 0.31 | 508 | 0.5 | 21.9 | 347 | 11.3 | 2.1 | 1.6 | Narrow |
| | 1.3 | 2.5 | | 434 | | | | | | | Blueshifted Hβ/[OIII] |
| | | | | | | 8.4 | 499 | 4.2 | | | Blueshifted Hα/[NII] |
| CRSS1605.6+2554 | 8.4 | 4.5 | −0.14 | 356 | – | 49.3 | 459 | 19.5 | – | – | Narrow |
| CRSS1705.3+6049 | – | 99.4 | | 311 | | | | | | | Narrow |
| **EMSS NLXG** | | | | | | | | | | |
| MS1252.4−0457 | 4.1 | 23.1 | 0.26 | 445 | 2.5 | 31.5 | 317 | 25.6 | 3.2 | 3.9 | Narrow* |
| | | 20.4 | | 686 | | | | | | | Blueshifted [OIII] |
| MS1334.6+0351 | 2.3 | 16.0 | 0.15 | 315 | 0.6 | 13.3 | 280 | 15.7 | 2.7 | 2.4 | Narrow |
| | 13.1 | | | 3669 | | 48.1 | 2768 | | | | Broad Hα/Hβ |
| MS1412.8+1320 | 7.3 | 9.9 | −0.35 | 314 | 0.8 | 28.1 | 163 | 10.2 | 4.2 | 5.2 | Narrow |
| | 20.3 | | | 2613 | | 80.9 | 1889 | | | | Broad Hα/Hβ |
| MS1414.8−1247 | 8.9 | 45.2 | 0.27 | 441 | 2.3 | 53.7 | 373 | 13.1 | 5.8 | 5.6 | Narrow |
| | | 10.0 | | 550 | | | | | | | Blueshifted [OIII] |
| | 25.9 | | | 2250 | | 89.5 | 1745 | | | | Broad Hα/Hβ |
| MS1555.1+4522 | 7.9 | 48.6 | 0.24 | 392 | 5.3 | 33.9 | 366 | 33.0 | 10.4 | 8.5 | Narrow |
| | | 15.7 | | 642 | | | | | | | Blueshifted [OIII] |
| MS1614.1+3239 | 1.3 | 16.7 | 0.20 | 423 | 1.4 | 3.4 | 363 | 8.9 | 3.0 | 1.8 | Narrow |
| | | | | | | 49.4 | 3725 | | | | Broad Hα |
| MS2044.1+7532 | 11.2 | 24.5 | 0.30 | 385 | 0.7 | 16.7 | 363 | 13.1 | 3.0 | 5.1 | Narrow |
| | | 13.1 | | 816 | | | | | | | Blueshifted [OIII] |
| | | | | | | 90.3 | 4142 | | | | Broad Hα |

∗ Hα/N[II] on atmospheric B band.
† Whittle (1985) asymmetry parameter, corresponding to the relative wavelength shift between the 10 percentile areas in the blue and red wings of the emission-line profile from the line centre.

asymmetries with the additional fitted components can be seen in figure 2, where we have plotted expanded spectra of the regions surrounding the major emission lines for a representative sample of 4 NLXG observed in this survey. The strengths of the additional [OIII]λ5007 components range from 25 per cent to 88 per cent of principal line, with velocity shifts between the two fitted components ranging from 210 km s$^{-1}$ to 653 km s$^{-1}$ (see table 3). We have confirmed that the summed equivalent width of both fitted components in these asymmetric lines is also good estimate (accurate to within 15 per cent) of the total equivalent width measured without line-fitting.

Similarly asymmetric [OIII]λ5007 profiles have also been seen in active galaxies by Whittle (1985). Whittle identified blue asymmetries in most of the AGN/HII regions he studied. If we use the same asymmetry parameter as defined by Whittle ($a(20\%)$), we find that the five NLXG in this analysis which require blueshifted components all have an asymmetry parameter $a(20\%) > 0.2$. In Whittle's sample approximately 40 per cent of the AGN had asymmetries with $a(20\%) > 0.2$, and so the results found here would appear to be consistent with Whittle's observations. Based on the Kolmogorov-Smirnoff (KS) test statistic, we find that the overall distribution of the measured [OIII]λ5007 line asymmetries (irrespective of their significance level) for the NLXG sample is consistent at the 95 per cent confidence level with that observed by Whittle (1985). Measured $a(20\%)$ values for all [OIII]λ5007 lines are given in table 2. The origin of these asymmetries is still a matter of some debate, but they are most likely to originate from wind-driven nuclear

6   B.J. Boyle et al.

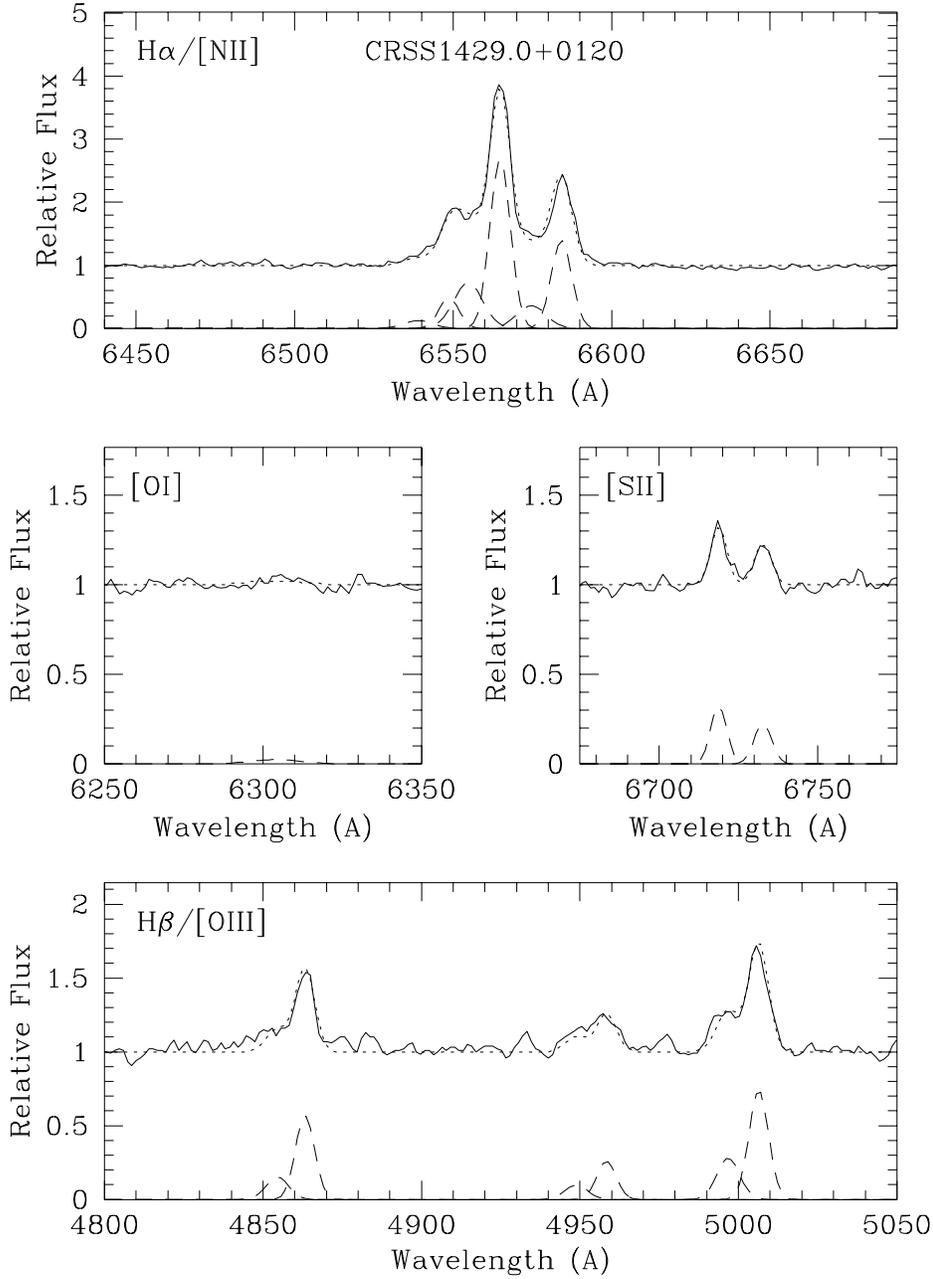

**Figure 2** Expanded spectra of the regions around the prominent emission lines in a respresentative sample of 4 NLXG observed in our sample. The spectra have been corrected to the rest frame and been divided through by a low-order polynomial fit to the continuum. The accepted fit is denoted by the short dashed line. The contribution of the individual emission lines is shown by the longer dashed lines. (a) CRSS 1429.0 + 0120: a starburst galaxy with asymmetric [OIII]/H$\beta$ and [NII]/H$\alpha$.

outflows in which dust preferentially obscures the emission from the far (red) side (Whittle 1985, Veilleux 1991).

For each spectrum, separate fits were then carried out for the following combination of lines over the wavelength intervals indicated: H$\beta$/[OIII]$\lambda\lambda 4959, 5007$ (4800Å – 5070Å); [OI]$\lambda 6300$ (6280Å – 6320Å); [NII]$\lambda 6549$/H$\alpha$/[NII]$\lambda 6584$ (6440Å – 6690Å); [SII] $\lambda\lambda 6717, 6734$ (6697Å – 6754Å). For each narrow emission line we tried two fits; a single narrow (FWHM $< 1000\,\mathrm{km\,s^{-1}}$) gaussian, and two narrow gaussians with an additional blueshifted component. For the Balmer lines we also tried a fit which comprised a narrow plus broad (FWHM $> 1000\,\mathrm{km\,s^{-1}}$) gaussian profile. To improve the robustness of the fits, we made every attempt to minimize the number of free parameters in each fit. As discussed above, the continuum-divided spectra first allowed us to fix the continuum at a constant value of 1. We also fixed the [NII]$\lambda 6734$:$\lambda 6717$ emission line ratio at 3.01:1 and the [OIII]$\lambda 5007$:$\lambda 4959$ emission line ra-tio at 2.88:1. Similarly, we fixed the relative rest wavelength ratios of the [NII], [SII] and [OIII] emission line pairs at 6549:6584, 6717:6734 and 4959:5007 respectively. Finally, for each group of narrow lines fitted simultaneously (i.e. Hβ/[OIII]λλ4959, 5007, [NII]λ6549/Hα/[NII]λ6584 and [SII] λλ6717, 6734), we adopted a single value for the FWHM of the narrow gaussian component fitted to all lines.

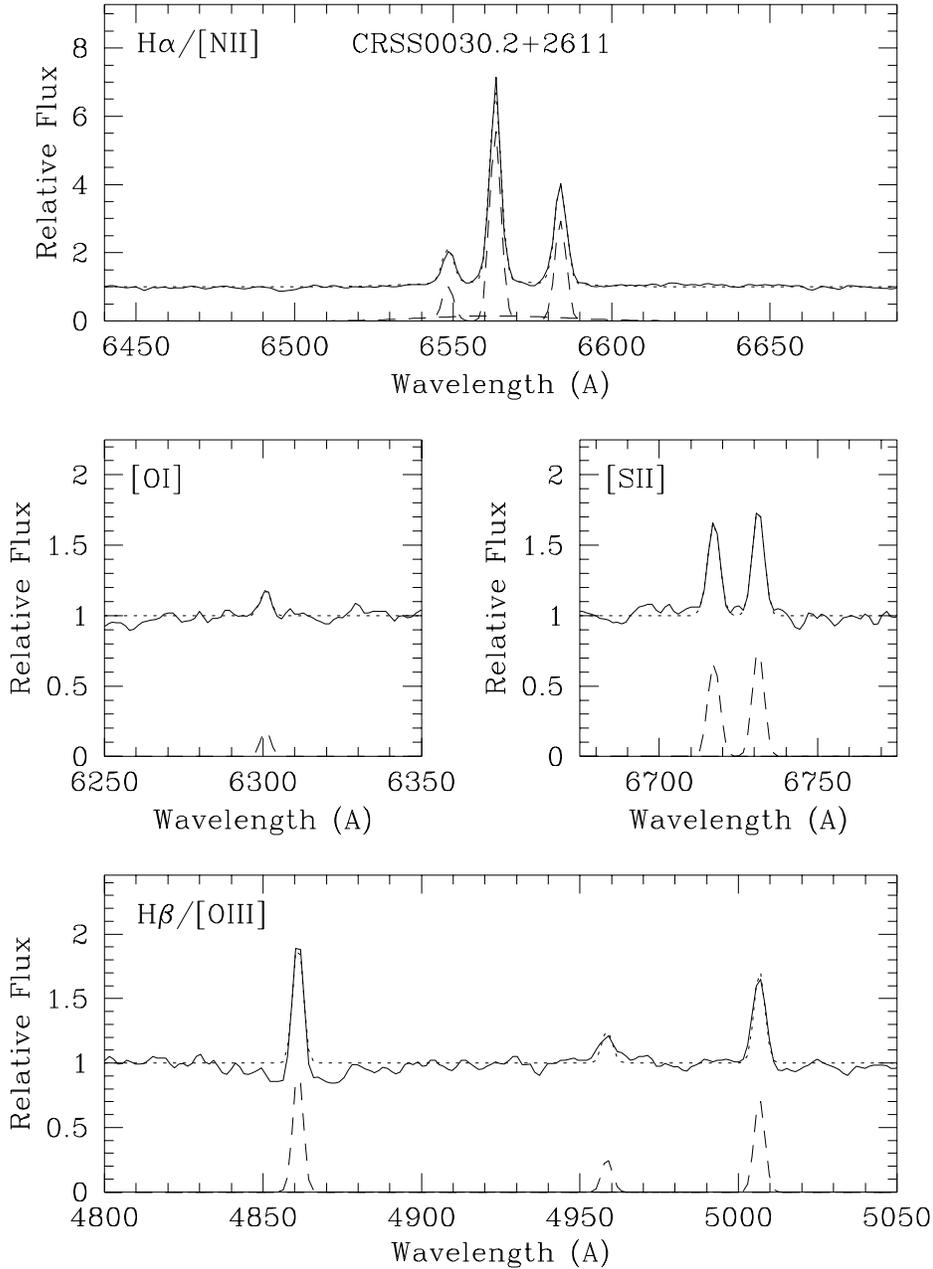

**Figure 2(b)** CRSS0030.2 + 2611: a starburst galaxy with weak broad Hα.

For each set of fits (narrow only, narrow plus blueshifted component, narrow plus broad) we again used the $F$-ratio test to discriminate between them, accepting the more complex fit (i.e. including the blueshifted or broad components) where it was preferred to the narrow-line-only fit at greater than 99 per cent confidence level.

## 3 RESULTS

Based on the fitting procedure outlined in the previous section, the measured rest-frame emission line equivalent widths and FWHM for each NLXG are listed in table 2. For each object the data is presented in two lines; the first provides the information for the narrow lines and the second lists the measured parameters for the second component (blueshifted line or broad emission line), if present. The identification of each component (narrow, blueshifted or broad) is given at the end of each line. Expanded spectra of the emission line regions in 4 reprensentative NLXG are plotted in figure 2. In each spectrum the total fit is shown by the short dashed line with the individual emission lines components represented by long dashed lines. The principal narrow emission line ratios, together with the velocity shift ($\Delta v$) of any blueshifted [OIII]$\lambda$5007 components fitted, are given in table 3. The emission line ratios quoted in table 3 (and used throughout the following discussion) are ratios of the total narrow emission-line equivalent widths (i.e. including both components of any 'double gaussian') and not intensity ratios. For the Balmer line ratios, the subscripts in table 3 refer to the ratio of the narrow (1) or broad (2) components. Similarly, the subscripts on the [OIII]$\lambda$5007 line ratios in table 3 correspond to the rest (1) and blueshifted (2) components respectively. For a power-law continuum $f_\nu \propto \nu^{-\alpha}$ it can be shown straightforwardly (see Boyle 1990) that the ratio of equivalent widths $W_1$ and $W_2$, measured at $\lambda_1$ and $\lambda_2$ respectively correspond to a ratio of intensities $I_1$ and $I_2$:

$$\frac{I_1}{I_2} = \frac{W_1}{W_2} \left(\frac{\lambda_2}{\lambda_1}\right)^{2-\alpha}$$

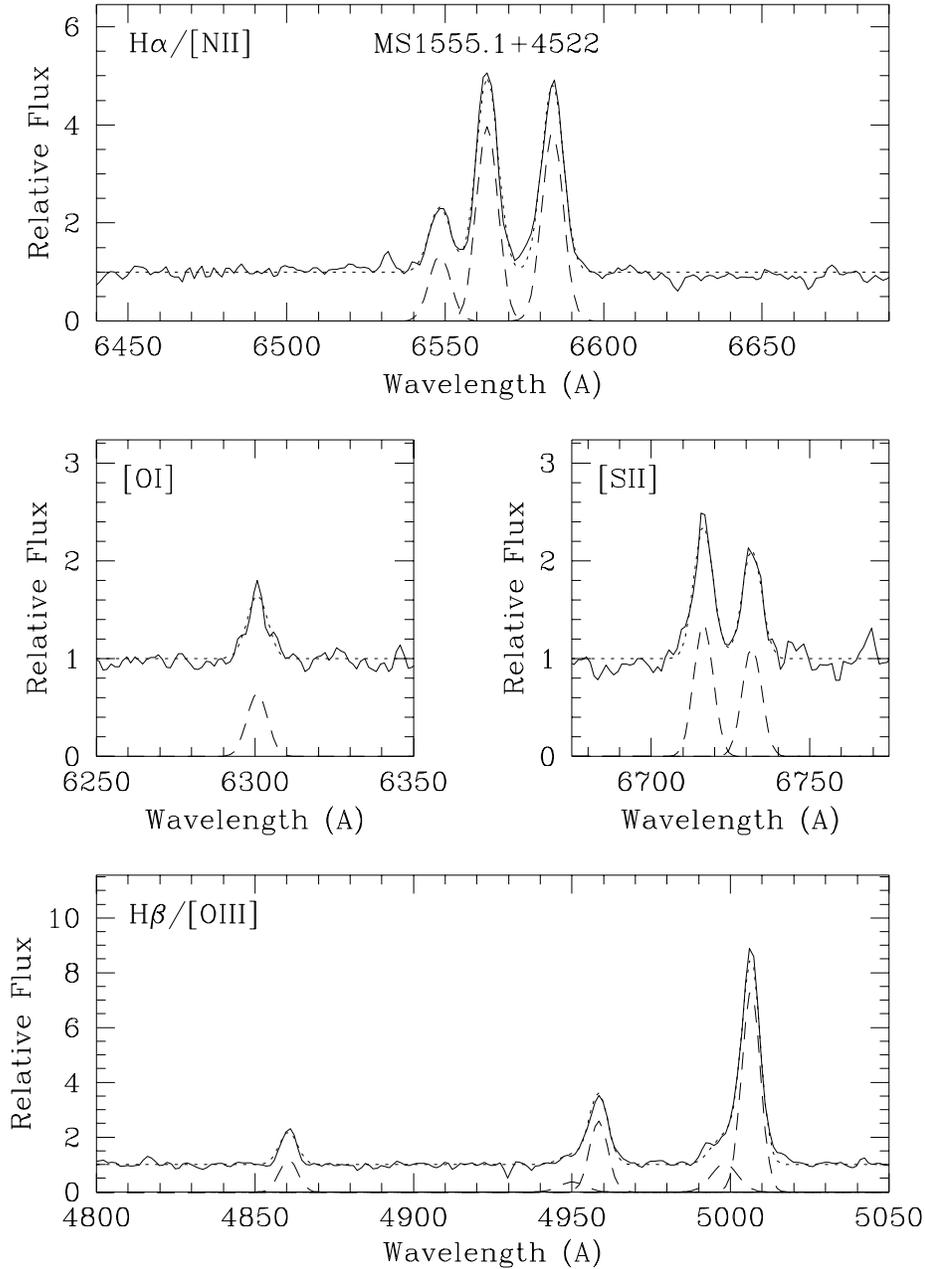

**Figure 2(c)** EMSS1555.1 + 4522: a Seyfert 2 galaxy with asymmetric [OIII].

With the exception of the H$\alpha$/H$\beta$ line ratio (discussed separately below), the correction from equivalent width ratio to intensity ratio is negligible for all line ratios quoted in table 3 (e.g. < 4 per cent in the [OIII]/H$\beta$ ratio) for most realistic values of the optical spectral index, $\alpha_{\rm opt} \sim 0.5$.

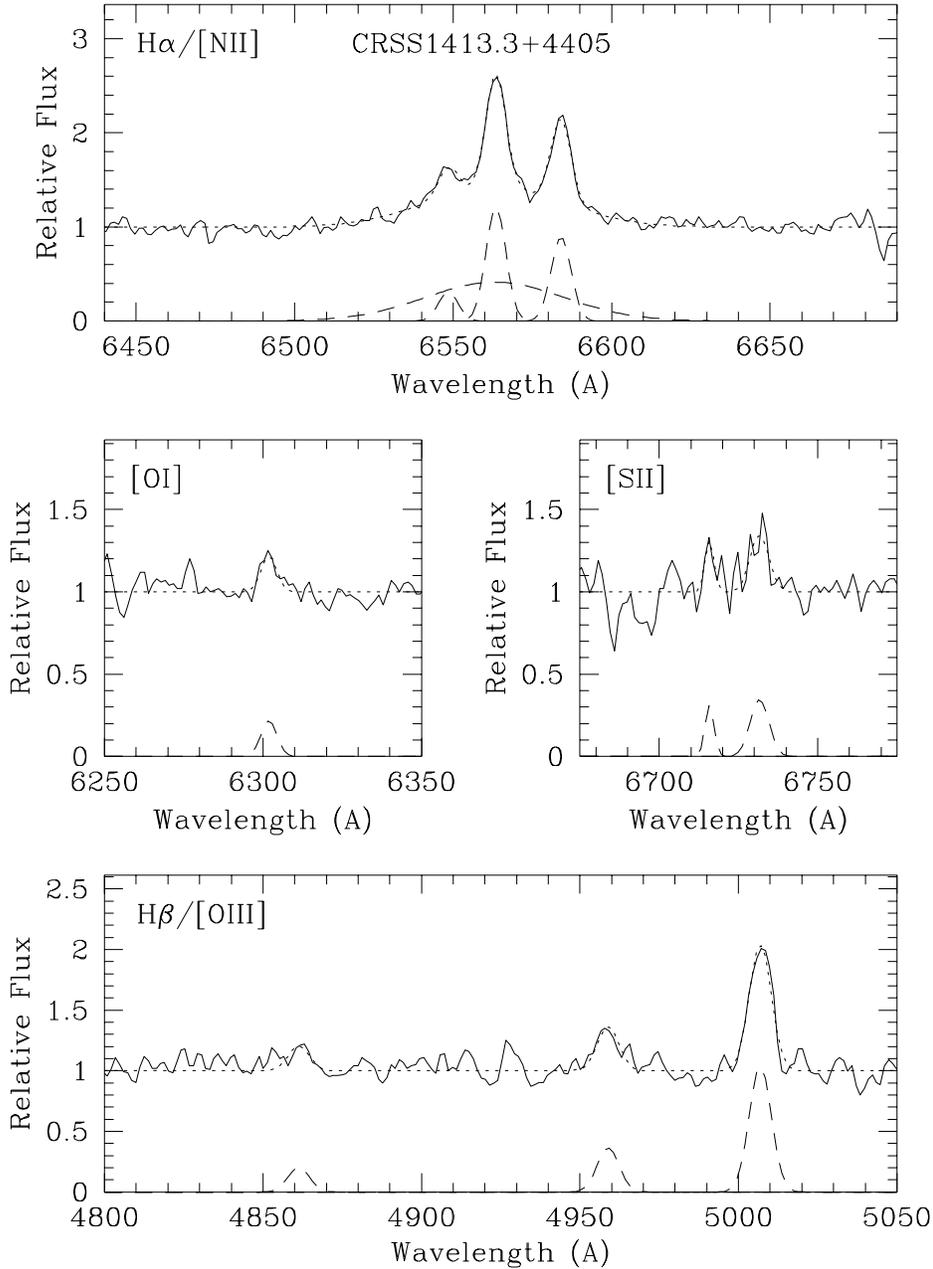

**Figure 2(d)** CRSS1413.3+4405: a Seyfert 1.5 galaxy.

In table 3 we also list the classification assigned to each NLXG on the basis of its position in the [NII]$\lambda$6584/H$\alpha$ v [OIII]$\lambda$5007/H$\beta$ diagram (see figure 3) using the scheme of Baldwin, Philips and Terlevich (1981). From figure 3, we can see that there is a good separation between the objects with HII-like spectra (i.e. starburst galaxies) and AGN-like spectra (Seyferts 1.5-2). The values of the other emission line ratios e.g. [OI]$\lambda$6300/H$\alpha$ and [SII]$\lambda$6717 + 6734/H$\alpha$ in each NLXG are also consistent with the classifications based on this diagram. For two NLXGs we have no observations of the H$\beta$/[OIII]$\lambda$5007 region. The low value of the [OI]$\lambda$6300/H$\alpha$ and [SII]$\lambda$6717 + 6734/H$\alpha$ ratios in CRSS0009.0+2041 mean that this object is likely to have an HII-like spectrum, whereas the much higher [OI]$\lambda$6300/H$\alpha$ ratio in CRSS0030.7+2629 implies an AGN-like spectrum (see Filippenko & Terlevich 1992). Although only the [OIII]$\lambda$5007 line is reliably detected in CRSS1705.3+6049 the upper limit of the equivalent width of the much weaker H$\beta$ suggests that this object is also likely to be an AGN, although this classification is still rather uncertain. For each object identified as an AGN, we further classified the object as a Seyfert 1.5, 1.8, 1.9 or 2, based on the relative strengths of the broad and narrow H$\alpha$ components using the approximate relation given by Netzer (1990): 1 + (Narrow/Total)$^{0.4}$.

**Table 3** Emission line ratios of NLXG sample

| Name | $\frac{[OIII]}{H\beta}$ | $\frac{[OI]}{H\alpha}$ | $\frac{[NII]}{H\alpha}$ | $\frac{[SII]}{H\alpha}$ | $\frac{H\alpha_1}{H\beta_1}$ | $\frac{[SII]_{6717}}{[SII]_{6734}}$ | $\frac{H\alpha_2}{H\alpha_1}$ | $\frac{H\beta_2}{H\beta_1}$ | $\frac{H\alpha_2}{H\beta_2}$ | $\frac{[OIII]_2}{[OIII]_1}$ | $\Delta v$ (km/s) | Classification |
|---|---|---|---|---|---|---|---|---|---|---|---|---|
| **CRSS NLXG** | | | | | | | | | | | | |
| CRSS0009.0+2041 |  | 0.02 | 0.67 | 0.20 |  | 1.21 | 0.29 |  |  |  |  | HII |
| CRSS0030.2+2611 | 0.82 | 0.03 | 0.53 | 0.26 | 6.73 | 0.91 | 0.31 |  |  |  |  | HII |
| CRSS0030.7+2616 |  | 0.08 | 1.28 |  |  |  |  |  |  |  |  | Sy 2 |
| CRSS1406.7+2838 | 0.61 | < 0.05 | 0.14 | 0.14 | 9.83 | 1.53 |  |  |  |  |  | HII |
| CRSS1412.5+4355 | 3.98 | 0.04 | 0.63 | 0.38 | 5.17 | 1.12 | 0.62 |  |  |  |  | Sy 1.8 |
| CRSS1413.3+4405 | 5.35 | 0.15 | 0.73 | 0.39 | 5.64 | 0.42 | 2.35 |  |  |  |  | Sy 1.5 |
| CRSS1415.0+4402 | 3.74 | 0.13 | 0.77 | 0.59 | 4.89 | 1.04 | 5.27 |  |  |  |  | Sy 1.5 |
| CRSS1429.0+0120 | 1.49 | 0.02 | 0.51 | 0.12 | 5.51 | 1.31 | 0.38 | 0.31 | 6.46 | 0.44 | 653 | HII |
| CRSS1605.6+2554 | 0.54 | < 0.02 | 0.40 | < 0.04 | 5.87 |  |  |  |  |  |  | HII |
| CRSS1705.3+6049 | > 10 |  |  |  |  |  |  |  |  |  |  | Sy 2? |
| **EMSS NLXG** | | | | | | | | | | | | |
| MS1252.4−0457 | 10.59 | 0.08 | 0.81 | 0.23 | 7.68 | 0.82 |  |  |  | 0.88 | 383 | Sy 2 |
| MS1334.6+0351 | 6.96 | 0.04 | 1.18 | 0.38 | 5.78 | 1.12 | 3.61 | 5.70 | 3.67 |  |  | Sy 1.5 |
| MS1412.8+1320 | 1.36 | 0.03 | 0.36 | 0.33 | 3.85 | 0.81 | 2.88 | 2.78 | 3.99 |  |  | HII |
| MS1414.8−1247 | 6.20 | 0.05 | 0.24 | 0.21 | 6.03 | 1.04 | 1.67 | 2.90 | 3.45 | 0.24 | 210 | HII |
| MS1555.1+4522 | 8.14 | 0.16 | 0.97 | 0.56 | 4.29 | 1.22 |  |  |  | 0.32 | 425 | Sy 2 |
| MS1614.1+3239 | 12.85 | 0.41 | 2.62 | 1.41 | 2.61 | 1.67 |  |  |  |  |  | Sy 1.5 |
| MS2044.1+7532 | 3.36 | 0.04 | 0.78 | 0.48 | 1.49 | 0.59 | 5.41 |  |  | 0.53 | 491 | Sy 1.5 |

Based on the observed emission-line ratios in the 17 NLXGs observed, we have identified 7 starburst galaxies and 10 AGN (including 5 Sy 1.5, 4 Sy 1.8-2 galaxies and one uncertain classification). We found no LINERS (see Heckman 1980) in the NLXG sample. These results are broadly in agreement with the results of Fruscione, Griffiths and MacKenty (1993), who found similar numbers of starburst galaxies and Seyfert galaxies amongst a similar sample of EMSS 'ambiguous' sources. Using the KS test, we were able to determine that there is no significant difference in the relative numbers of starburst galaxies/AGN found in the CRSS and EMSS samples.

Indeed, the AGN and the starbursts cannot be distinguished in the present small sample by X-ray luminosity, redshift, optical magnitude, presence of a broad component, or line asymmetry. There is also no evidence for any X-ray luminosity dependence in the ratio of starburst galaxies to AGN. The X-ray to optical ratio ($\alpha_{OX}$) might be able to discriminate, since nearby starbursts are relatively X-ray faint (Fabbiano 1989), but the large galaxy contribution to the optical magnitudes prevents us measuring this ratio in a meaningful fashion. High resolution imaging is needed.

We confirm the provisional classification of Seyfert-like galaxies and starburst-like galaxies assigned to all but one of the CRSS and EMSS NLXG in Paper I. The exception, CRSS1429.0+0120, is now classified as a starburst galaxy (due to an improved measurement of the emssion line ratios), although it is one of the objects closest to the transition region between HII and AGN in Fig. 3. One further NLXG (MS2044.1+7532), originally classified as a borderline HII/Sy2 object, is now identified as a Seyfert 1.5 galaxy. Indeed, the unambiguous detection of broad H$\alpha$ in many NLXG allows us to refine the classification of Seyfert galaxies presented in Paper I (originally referred to simply as Sy2-like galaxies). CRSS0030.2+2611, CRSS1705.3+6049, MS1252.4−0457 and MS1555.1+4522 retain their Seyfert 2 classification, while CRSS1412.5+4355 is re-classified as a Seyfert 1.8 and CRSS1413.3+4405, CRSS1415.0+4402, MS1334.6+0351 and MS1614.1+3239 are re-classified as Seyfert 1.5. We also note that the [OIII]$\lambda$5007/H$\beta$ and [NII]$\lambda$6584/H$\alpha$ emission line ratios quoted in Paper I took no account of any broad component in H$\beta$ or H$\alpha$ which may have been present and will therefore display systematic differences to the emission line ratios presented in this paper. Any other discrepancies between the quoted emission line ratios will be due to the measurement process; the method adopted in this paper being more reliable.

The FWHM of the narrow H$\alpha$ lines lies in the range $170 < \text{FWHM} < 460 \text{ km s}^{-1}$, with no significant difference between the distribution of FWHM for the starburst galaxies and AGN samples. Nine NLXG were found to exhibit broad H$\alpha$ components, including 6 of the 9 objects classified as AGN on the basis of their emission line ratios. Three of the starburst galaxies (CRSS0030.2+2611, MS1412.8+1320, MS1414.8−1247) also exhibit broad components, with FWHM ranging from $1700 \text{ km s}^{-1}$ to $2400 \text{ km s}^{-1}$. Broad H$\alpha$ profiles (up to FWHM = $3500 \text{ km s}^{-1}$) have previously been detected in HII regions in starburst galaxies (e.g. NGC2363, Roy et al. 1992, Gonzalez-Delgado et al. 1994), although their origin is uncertain (stellar winds, supernovae remnants, superbubbles). The equivalent widths of the broad H$\alpha$ components observed in this sample of starburst galaxies are also roughly consistent with the range observed in NGC2363 (few Å - 40Å). However, we cannot rule out the possibilty that the broad emission is due to a 'mini-QSO' embedded in the starburst galaxy.

The broad H$\alpha$ emission lines in the AGN sample have a mean value of $3900 \pm 1900 \text{ km s}^{-1}$. Broad H$\beta$ was only conclusively detected in 1 AGN. This has important consequences for the classification of such objects from spectra with limited wavelength coverage, particularly in the case when the region around Hα is not observed. For NLXGs with $z > 0.25$, this will frequently be the case.

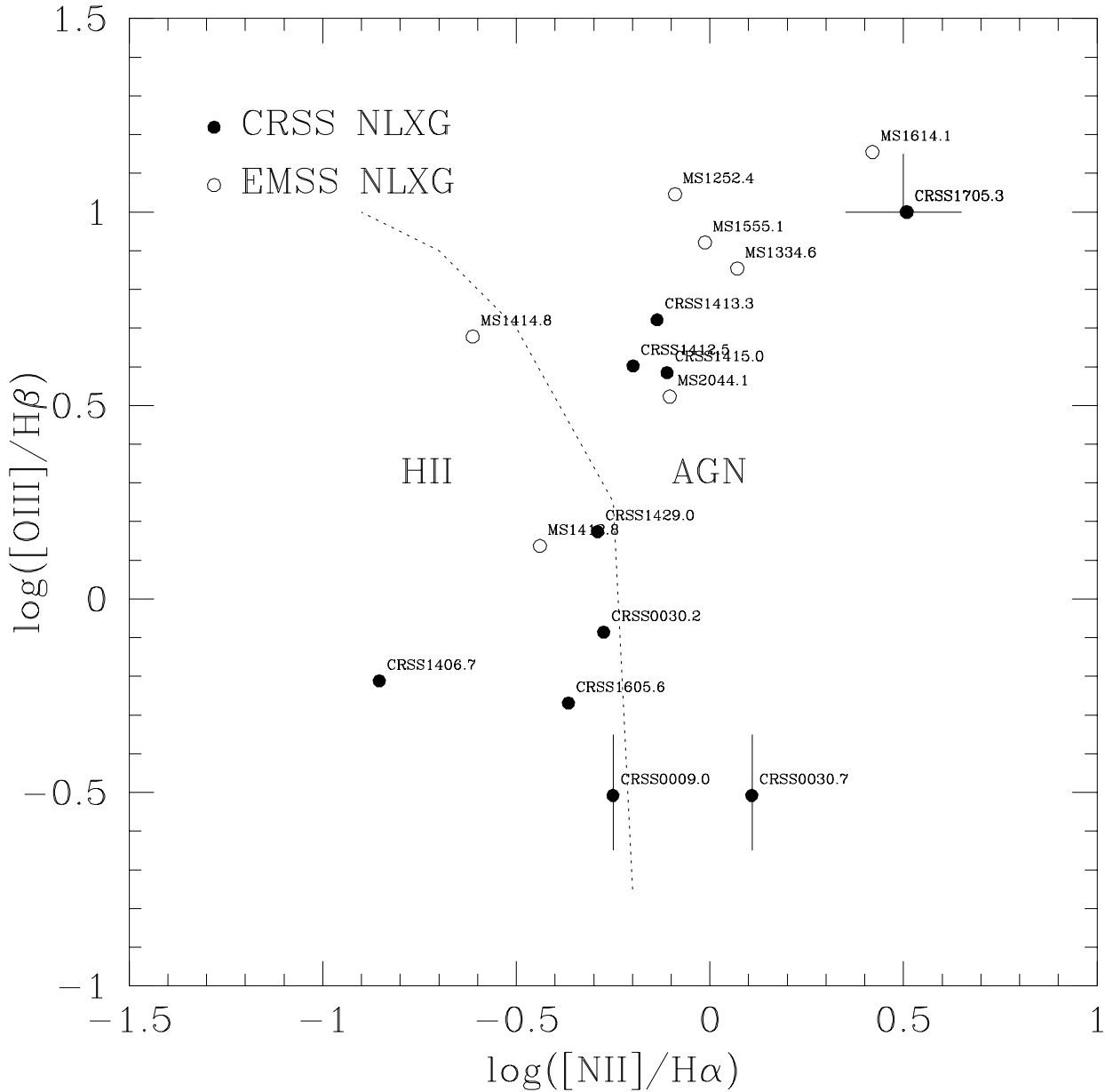

**Figure 3** [NII]λ6584/Hα – [OIII]λ5007/Hβ emission-line ratio diagram for the NLXGs observed in this paper. CRSS objects are indicated by the filled circles and EMSS objects by the open circles. The division between AGN-like and HII-like spectra (dotted line) is based on the criterion of Baldwin, Philips & Terlevich (1981).

The Hα/Hβ equivalent width ratios for the sample range from 1.5 to 9.8, with a mean of 5.4. The mean values for the Seyfert and starburst galaxy samples are 4.6 (range 1.5 to 7.7) and 6.3 (range 3.8 to 9.8) respectively. In order to derive Hα/Hβ intensity ratios, we have multiplied these equivalent width ratios by 0.64, i.e. assuming a optical spectral index $\alpha_{opt} = 0.5$ (see above). Note that this factor is relatively insensitive to spectral index, only changing from 0.58 to 0.78 even over the wide range in spectral indices, $0.2 < \alpha_{opt} < 1.2$ observed by Francis et al. (1992). The mean Seyfert galaxy Hα/Hβ intensity ratio derived in this manner is 2.9, consistent with the value predicted from photoionisation models (Netzer 1990). For the starburst galaxies, the mean value is 4.0, although it reduces to 3.6 if the anomalously high Hα/Hβ ratio measured from the low signal-to-noise spectrum of CRSS1406.7+2838 is excluded. This value is slightly higher than the predicted range in the Hα/Hβ intensity ratios for HII regions: $2.8 < Hα/Hβ < 3.0$ (Aller 1974). However, given the typical uncertainties in the derivation of these narrow-line ratios (dominated by the

12  B.J. Boyle et al.

$\sim 15$ per cent uncertainty in the equivalent width measurement of each individual line), this is not a significant discrepancy. The lack of significant reddening from the observed H$\alpha$/H$\beta$ ratios is also consistent with the results from the X-ray spectral analysis (Ciliegi et al. 1995) in which none of the NLXG X-ray spectra (with the exception of CRSS1412.5+4355) show any evidence for any intrinsic absorption due to neutral hydrogen in excess of the galactic value (although few NLXG have sufficient X-ray counts to permit a detailed spectral fit). The Seyfert 1.8 galaxy CRSS1412.5+4355 has an intrinsic neutral hydrogen X-ray column density $N_H = 2.5 \pm 1.0 \times 10^{20}$ cm$^{-2}$, corresponding (for galactic gas-to-dust ratios) to a small visual extinction of $A_V = 0.14$ mag (Zombeck 1990). This small amount of extinction is consistent with the mild amount of reddening implied by the intensity ratio H$\alpha$/H$\beta$ = 3.3 derived for CRSS1412.5+4355. Thus, while individual NLXG may exhibit some reddening, it would appear that significant obscuration is not a general feature of either the starburst or Seyfert population in this sample.

As demonstrated in Paper I, the NLXG sample comprises between 15–35 per cent of the soft (2 keV) X-ray background. With an approximate ratio of 10:7 AGN:starbursts identified in this paper, this suggests that the approximate contributions of the two populations also lie in the approximate ratio 10:7 per cent. However, the numbers of NLXGs identified are still small, and we can not rule out equal contributions from both classes of object. Given the composition of the NLXG sample, it is not surprising that the rate of cosmological evolution derived in Paper I, $L_X \propto (1+z)^{2.6\pm 1}$, is so similar to that of QSOs ($L_X \propto (1+z)^{3.0\pm 0.2}$). Unified models of AGN (in which the appearance of an object as a Seyfert 1 or 2 is merely dependent on viewing angle) naturally imply that Seyfert 2s (or similar types) must evolve at the same rate as Seyfert 1s/QSOs. In addition, it is also known that starburst galaxies also undergo a rate of cosmological evolution in infra-red luminosity $L_{IR} \propto (1+z)^{3.0\pm 1.0}$ (Saunders et al. 1990) which is consistent with that of QSOs in the optical and X-ray regimes (see Boyle 1993).

## 4 CONCLUSIONS

We have obtained intermediate-resolution spectra of 17 NLXG identified from the CRSS and EMSS samples. Based on their emission line ratios, we estimate that the sample contains 7 starburst galaxies and 10 Seyfert galaxies. Six of the Seyfert galaxies show evidence for broad H$\alpha$ emission, although only one conclusively exhibits broad H$\beta$ emission. The Seyfert types range from Seyfert 1.5 to 2. In addition, 3 of the starburst galaxies exhibit evidence for weak broad ($\sim 2000$ km s$^{-1}$) H$\alpha$ emission. Thus, the NLXG sample as originally identified in Paper I appears to be a heterogeneous mix of Seyfert and starbursts. Only the line ratios distinguish the two classes. In all other characteristics they share similar properties ($z$, optical magnitude, $L_X$, and line asymmetry distributions). If the two classes are powered by different processes, this is surprising. Further discriminants need to be searched for in larger samples. Both classes contribute approximately equally to the 2 keV X-ray background at a level of between 7 and 17 per cent.

## ACKNOWLEDGEMENTS

RGM acknowledges the receipt of a Royal Society University Research Fellowship. We are indebted to Dr Mike Irwin for observing the two spectra obtained during spectroscopic service time on the William Herschel Telescope. We also would also like to thank Dr Gerard Kriss for providing help with the SPECFIT routine. BJB acknowledges the support and hospitality of the Smithsonian Astrophysical Observatory. The X-ray data was obtained from the Leicester and Goddard *ROSAT* archives. This work was partially supported by NASA grants NAGW-2201 (LTSA) and NAS5-30934 (RSDC). The optical spectra were obtained at the William Herschel Telescope at the Observatory of the Roque de los Muchachos operated by the Royal Greenwich Observatory.

## REFERENCES

Aller L.H. 1974, in Physics of Thermal Gaseous Nebulae, Reidel, Dordrecht, p.77
Baldwin J. A., Philips M. M., Terlevich R. J., 1981, PASP, 93, 5
Böller Th., Meurs E.J.A., Brinkman W., Fink H., Zimmermann U., Adorf H.-M., 1992, A&A, 261, 57
Boyle B. J., 1990, MNRAS, 243, 231
Boyle B. J. 1993, in Shull, J.M., Thronson, H.A., eds, The Environment and Evolution of Galaxies, Kluwer, Dordrecht, p.433
Boyle B. J., Shanks T., Georgantopoulos I. G., Stewart G. C., Griffiths R. E., 1994, MNRAS, 271, 639
Boyle B. J., McMahon R. G., Wilkes B. J., Elvis M., 1995, MNRAS, 272, 462 (Paper I)
Ciliegi P., Elvis M., Wilkes B. J., Boyle B. J., McMahon R. G. 1995, MNRAS, submitted
Fabbiano G., 1989, Ann. Rev. A&A., 27, 87
Fabian A. C., Barcons X., 1992, ARAA, 30, 329
Filippenko A. V., Terlevich R.J., 1992, ApJ, 397, L79
Francis P. J., Hewett P. C., Foltz C. B., Chaffee F. H., 1992, ApJ, 398, 476
Fruscione A., Griffiths R. E., Mackenty J. W, 1993, in Crincarini,D. et al. eds. Observational Cosmology, ASP Conf. Ser. 51, Astron. Soc. Pacif., San Francisco, p296
Georgantopoulos I. G., Stewart G. C., Shanks T., Griffiths R. E., Boyle B. J., 1993, MNRAS, 262, 619
Georgantopoulos I. G., Stewart G. C., Shanks T., Griffiths R. E., Boyle B. J., 1995, MNRAS, in press
Gonzalez-Delgado M. et al., 1994, ApJ, 437, 239
Griffiths R. E., Padovani P., 1990, ApJ, 360, 483
Heckman T. M. 1980, A&A, 87, 152
Irwin M. J., McMahon R.G., Maddox S.J., 1994, Spectrum, 2, 14
Mood R., Graybill R., 1963, Introduction to the theory of statisitics, (2nd edition), p231
Netzer H., 1990, in Active Galactic Nuclei, Saas-Fee Advanced Course 20, Springer, Berlin, p57
Roche N., Shanks T., Georgantopoulos I., Stewart G. C., Boyle B.J., Griffiths R. E., 1995, MNRAS, 275, L15
Roy J-R., Boulesteix J., Joncas G.,Grundseth B., 1992, ApJ 386, 498
Saunders W., Rowan-Robinson, M., Lawrence A., Efstathiou G., Kaiser N., Ellis R. S., Frenk C. S. 1990, MNRAS, 242, 318
Stocke J. T., Morris S. L., Gioia I. M., Maccacaro T., Schild R., Wolter A., Fleming T. A., Henry J. P., 1991, ApJS, 76, 813
Whittle M. 1985, MNRAS, 213, 1

Veilleux S., 1991, ApJS, 75, 383
Zombeck M. 1990, in Handbook of Space Astronomy and
    Astrophysics, Cambridge University Press, p.103